\documentclass[aps,pre,twocolumn]{revtex4-1}

\usepackage{amsmath,amssymb,mathrsfs,bm}
\usepackage{color,graphicx,subfigure}

\newcommand{\bA}{\mathbf{A}}
\newcommand{\ba}{\mathbf{a}}
\newcommand{\Bref}{\mathscr{B}_0}
\newcommand{\Brel}{\mathscr{B}_\text{rel}}

\newcommand{\bB}{\mathbf{B}}

\newcommand{\bBe}{\mathbf{B}_{\mathrm{e}}}
\newcommand{\bBeC}{\bBe^{\!\blacktriangledown}}
\newcommand{\bbD}{\mathbb{D}}
\newcommand{\bbI}{\mathbb{I}}
\newcommand{\bbT}{\mathbb{T}}
\newcommand{\bD}{\mathbf{D}}
\newcommand{\bE}{\mathbf{E}}
\newcommand{\be}{\mathbf{e}}
\newcommand{\bF}{\mathbf{F}}
\newcommand{\bFe}{\mathbf{F}_{\!\mathsf{e}}}

\newcommand{\bG}{\mathbf{G}}
\newcommand{\bH}{\mathbf{H}}
\newcommand{\bI}{\mathbf{I}}
\newcommand{\bk}{\bm{k}}
\newcommand{\bkc}{\mathbf{\hat{k}}}
\newcommand{\bPsi}{\bm{\Psi}}
\newcommand{\bPsiC}{\bPsi^{\!\blacktriangledown}}
\newcommand{\bl}{\bm{\ell}}
\newcommand{\bL}{\mathbf{L}}
\newcommand{\bM}{\mathbf{M}}
\newcommand{\bn}{\mathbf{n}}
\newcommand{\bno}{\mathbf{\mathring{n}}}
\newcommand{\R}{\mathbb{R}}
\newcommand{\bt}{\mathbf{t}}
\newcommand{\bT}{\mathbf{T}}
\newcommand{\bW}{\mathbf{W}}
\newcommand{\bu}{\mathbf{u}}
\newcommand{\bv}{\mathbf{v}}
\newcommand{\bx}{\mathbf{x}}
\newcommand{\bX}{\mathbf{X}}

\newcommand{\eps}{\varepsilon}
\newcommand{\piso}{p_{\text{iso}}}
\newcommand{\si}{\sigma}
\newcommand{\sio}{\sigma_{0}}
\newcommand{\Sym}{\mathit{Sym}}
\newcommand{\taur}{\tau_{\text{rel}}}
\newcommand{\taus}{\tau_{\text{s}}}
\newcommand{\taud}{\tau_{\text{d}}}
\newcommand{\taudef}{\tau_{\text{def}}}

\newcommand{\Ldot}[1]{\overset{\bm{.}}{#1}}
\newcommand{\vs}{v_{\text{s}}}
\newcommand{\ang}{\Theta}
\DeclareMathOperator{\tp}{\otimes}

\DeclareMathOperator{\kpp}{\widehat{\boxtimes}}
\DeclareMathOperator{\sym}{sym}
\DeclareMathOperator{\skw}{skw}
\DeclareMathOperator{\tr}{tr}
\DeclareMathOperator{\er}{e}
\DeclareMathOperator{\dev}{dev}
\DeclareMathOperator{\divr}{div}
\DeclareMathOperator{\re}{Re}

\newcommand{\D}[2]{\frac{\partial #1}{\partial #2}}

\newcommand{\bna}{\boldsymbol{\nabla}}
\begin{document}

\title{Viscoelastic nematodynamics}
\author{Stefano S.\ Turzi}
\affiliation{\mbox{Dipartimento di Matematica, Politecnico di Milano, Piazza Leonardo da Vinci 32, 20133 Milano, Italy}}

\date{\today}

\begin{abstract}
Nematic liquid crystals exhibit both crystal-like and fluid-like features. In particular, the propagation of an acoustic wave shows an unexpected occurrence of some of the solid-like features at the hydrodynamic level, namely, the frequency-dependent anisotropy of sound velocity and acoustic attenuation. The non-Newtonian behavior of nematics also emerges from the frequency-dependent viscosity coefficients. To account for these phenomena, we put forward a viscoelastic model of nematic liquid crystals, and we extend our previous theory to fully include the combined effects of compressibility, anisotropic elasticity and dynamic relaxation, at any shear rate. The low-frequency limit agrees with the compressible Ericksen-Leslie theory, while at intermediate frequencies the model correctly captures the relaxation mechanisms underlying finite shear and bulk elastic moduli. We show that there are only four relaxation times allowed by the uniaxial symmetry.
\end{abstract}

\maketitle

\section{Introduction}
\label{sec:introduction}
For nematics made of small molecules, the viscoelastic effects and the nonlinear effects are usually believed to be not important. Hence, the flow of nematics can be described by the Ericksen-Leslie theory\cite{95dgpr}, where the dissipative part of the stress tensor depends linearly on the strain rate. In this respect, nematic liquid crystals (NLCs) are non-Newtonian fluids only because of their anisotropic dissipation, while they show no static elastic anisotropy, no shear stiffness and their viscous coefficients do not depend on frequency. The dissipation function is thus quadratic in the velocity gradient, like in ordinary fluids. Elasticity appears only when director distortions are taken into account and the free energy is then supplemented with Frank's elastic energy.

However, this description is correct only at low-frequencies, i.e., for small strain rates. Actual soft matters are usually viscoelastic: they have both viscosity and bulk elasticity, and the relation between the stress and strain (or strain rate) is nonlinear. Viscoelastic features emerge also in nematic liquid crystals at ultrasonic frequencies. For instance, the anisotropy of sound speed and attenuation and their frequency dependence is often described in terms of an elastic material response and relaxation dynamics \cite{72muls,75Miyano,95gr,95kr,12kkkk}. Structural relaxation processes are also explicitly mentioned in order justify the frequency dependence of the viscosity coefficients \cite{72wetsel,73Jahnig,73eden,74kemp,75monroe,82blandin}. Recent papers even report the measurement of a viscoelastic response in low molecular weight liquid crystals, either in the nematic or the isotropic phase, when they are subjected to low-frequency mechanical sinusoidal deformations \cite{15Oswald,13kahl}.

While this ideas seem to be in good agreement with experiments they do not fit well with the existing hydrodynamic theories\cite{88sellers}. A comprehensive description of these phenomena along these lines of thoughts has only recently appeared \cite{14bdt,15tur,16bdt}. 
The theory put forth in Refs.\cite{14bdt,15tur,16bdt} has the advantage to enable a smooth transition from liquidlike to solidlike response. It is characterized by an anisotropic neo-Hookean contribution to the strain energy and an evolution of the relaxed (shear-stress free) configuration. In the two earlier papers \cite{14bdt,15tur}, we have constructed a simplified theory for (slightly) compressible NLCs and applied it, with fair success, to explain quantitatively the anisotropy of sound velocity \cite{72muls} and sound attenuation \cite{70lord} in $N$-(4-methoxybenzylidene)-4-butylaniline (MBBA) over the range 2--14 MHz. In a subsequent paper \cite{16bdt}, we have refined our theory to include the anisotropy of the dissipation tensor. The low-frequency limit of our theory reproduces the classical incompressible Ericksen-Leslie theory, but delivers in addition a new Parodi-like relation among the viscosity coefficients. 

Although it has long been recognized that sound propagation is affected by the orientational order\cite{70lord,72muls}, the study of acoustic phenomena in NLCs is still an active field of research \cite{Kapustina2004,Kozhevnikov2005,Kapustina2008,Kozhevnikov2010,Kapustina2014}. To some extent, this is due to the potential theoretical implications as there is not a broad agreement about the theoretical explanation of the interaction of sound and nematic order. Some time ago, a theory of sound propagation has been proposed by Selinger and co-workers to explain their experimental results \cite{02seli,03seli,04seli,05seli}. They postulated that the sound-speed anisotropy is due to a direct coupling between the nematic director and the density gradient. Later, this idea has been refined by Virga \cite{09virga} who has developed a thorough theory of anisotropic Korteweg (or second-gradient) fluids. However, here we do not share the point of view reported in \cite{02seli,09virga} and we have expressed our concerns 
in Ref.\cite{14bdt}. Furthermore, the second-gradient theory does not seem to predict the correct frequency dependence of the 
sound speed \cite{Kozhevnikov2005}.

In this paper, we reconsider and extend the model put forth in \cite{14bdt,15tur,16bdt}, to include the combined effects of (i) compressibility, (ii) an \emph{anisotropic neo-Hookean} contribution to the strain energy, and (iii) an \emph{anisotropic} gradient flow dynamics for the relaxed configuration. The acoustic approximation allows us to show that there are only four possible relaxation times. We calculate how the sound speed and the acoustic attenuation depend on these characteristic times and on the angle between the director and the direction of propagation. The weak-flow approximation yields the explicit dependence of the viscosity coefficients (including the bulk viscosities) on the parameters of the model.

\section{Dynamics}
\label{sec:dynamics}
We now review the main features of the theory presented in \cite{16bdt}. We revise it and extend it to take explicitly into account compressibility and the full spectrum of possible shear rates. Furthermore, we analyze more carefully the relaxation modes allowed by symmetry. 

The first key idea is to decouple the contribution of elasticity and dissipation by writing a multiplicative decomposition of the deformation gradient $\bF$ (with respect to an arbitrarily chosen reference configuration, $\Bref$)
\begin{equation} \label{eq:split}
\bF\!=\!\bFe\bG\,.
\end{equation}
The \emph{effective deformation tensor}, $\bFe$, will measure the \emph{elastic response} of the NLC, from an \emph{evolving equilibrium configuration}, $\Brel$. The \emph{relaxing deformation tensor}, $\bG$, determines how this configuration locally departs from the reference configuration. Since the elastic response is determined by $\bFe$, only the effective deformation appears explicitly in the strain energy. By contrast, energy  dissipation (entropy production) is only associated with the evolution of $\bG$.  

\begin{figure}[t]
\centering
\includegraphics[width=0.4\textwidth]{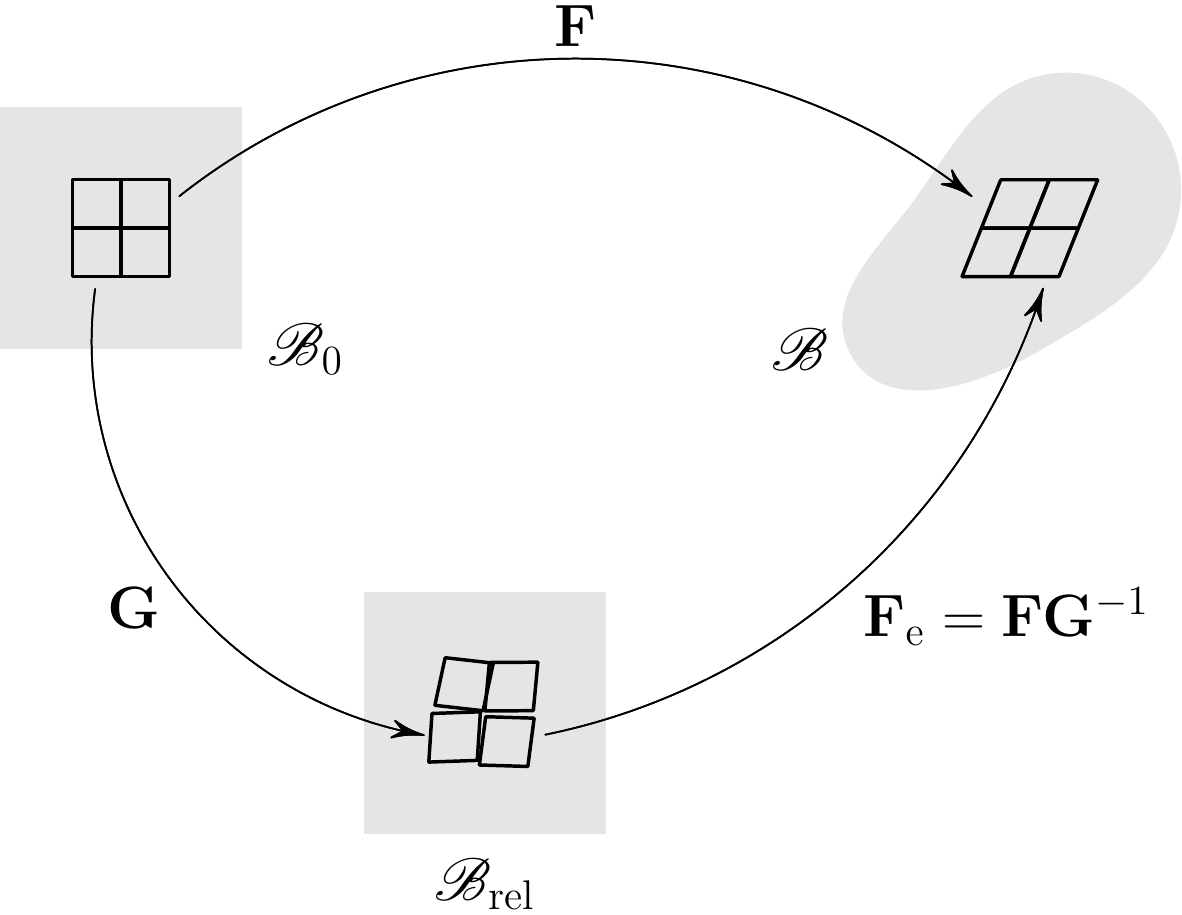}
\caption{Decomposition of the deformation gradient in its relaxing and effective parts.}
\label{fig:patatoidi}
\end{figure}
In order to account for anisotropic elasticity, we introduce the following compressible anisotropic neo-Hookean \emph{strain energy density} per unit mass 
\begin{equation}
\begin{split}
\si(\rho,\bBe,\bn) & = \sio(\rho) + \tfrac{1}{2}\mu(\rho) \Big(\tr \big(\bPsi^{-1}\bBe - \bI \big) \\
& - \log\det\big(\bPsi^{-1}\bBe \big)\Big),
\end{split}
\label{eq:StrainEnergy}
\end{equation}
where $\rho$ is the density, $\rho\mu(\rho)$ is the shear modulus, $\bBe=\bFe\bFe^\top$ is the \emph{effective} left Cauchy-Green deformation tensor and $\bI$ is the identity tensor. The isotropic contribution $\sio(\rho)$ penalizes density (and volume) variations and it is assumed to dominate the elastic energy. This term depends on $\rho$ and is thus not affected by stress relaxation. By contrast, the second term in \eqref{eq:StrainEnergy} depends on the evolution of $\Brel$. It couples the elastic properties of the material with the nematic director and it is minimum if and only if  $\bBe$ coincides with the \emph{energetic shape tensor}
\begin{equation}
\bPsi(\rho,\bn) = a(\rho)^2 (\bn \tp \bn) + a(\rho)^{-1}\big(\bI - \bn \tp \bn \big),
\label{eq:shape}
\end{equation}
where $\bn$ is the nematic director and $a(\rho)>0$ is a density-dependent aspect ratio. The shape tensor $\bPsi$ is symmetric, positive definite, and with unit determinant. If nematic distortions have to be included into the model, Eq.\eqref{eq:StrainEnergy} must be supplemented with the classical Oseen-Frank free-energy density \cite{95dgpr}. For simplicity, we assume that the director field, $\bn$, is uniform and fixed by some
external action.

A standard procedure \cite{14bdt,15tur,16bdt} then yields the Cauchy stress tensor
\begin{equation}
\bT = -p(\rho,\bBe,\bn) \,\bI + \rho \mu(\rho) \big(\bPsi^{-1}\bBe - \bI \big) ,
\label{eq:stress}
\end{equation}
where the pressure-like function is
\begin{equation}
\begin{split}
p & = \rho^2 \Big[\D{\sio}{\rho} + \tfrac{1}{2}\mu'(\rho)\Big(\tr \big(\bPsi^{-1}\bBe - \bI\big) \\
& \quad - \log\det\big(\bPsi^{-1}\bBe\big) \Big) \\
& \quad -\mu (\rho ) \frac{3 a'(\rho)}{2a(\rho)} \Big(\bn\tp\bn-\frac{1}{3}\bI\Big) \cdot \bPsi^{-1}\bBe \Big].
\end{split}
\label{eq:pressure}
\end{equation}
As expected, when $\bBe\!=\!\bPsi$, the shear stress vanishes and $\bT$ is spherical. 

The dynamics of the liquid crystal is governed by the balances of mass, momentum and angular momentum at each point of the system. Given our assumption on the director field, the macroscopic motion of the NLC is completely described by the balance of mass and the balance of momentum. In the absence of body forces and body couples these two equations are
\begin{align}
\dot{\rho} + \rho\divr \bv = 0, \qquad \rho \dot{\bv} = \divr \bT .
\label{eq:balances}
\end{align}
Here, a superimposed dot indicates the material time derivative and $\bv$ is the velocity field.

However, a full description of the dynamics requires an evolution equation for the relaxing deformation\cite{02DiCarlo,04Rajagopal}. Any relaxation dynamics necessarily obeys a dissipation inequality, ensuring a non-negative entropy production. It is convenient to introduce the \emph{co-deformational derivative}
\begin{equation}
\bBeC=\dot{\bB}_{\mathsf{e}}-(\bna\bv)\,\bBe-\bBe(\bna\bv)^{\!\top},
\end{equation}
where $\bna\bv\!=\!\dot{\bF}\bF^{-1}$ is the spatial velocity gradient. The quantity $\bBeC$ can be interpreted as the time rate of change of $\bBe$ relative to a convected coordinate system that moves and deforms with the flowing liquid crystal. In this respect, $\bBeC$ vanishes whenever the deformations are purely elastic and there is no evolution of the natural configuration (the system does not dissipate). The ``driving force'' for the evolution of $\Brel$ is $\D{\si}{\bBe}$. In this notation, the dissipation inequality for an isothermal process simply reads \cite{16bdt}
\begin{equation}
-\D{\si}{\bBe}\cdot \bBeC \geq 0.
\label{eq:ClDuin}
\end{equation}
When Eq.\eqref{eq:StrainEnergy} is substituted into \eqref{eq:ClDuin} this inequality takes the form
\begin{equation}
\left(\bPsi^{-1} - \bBe^{-1}\right) \cdot\bBeC  \leq 0,
\label{eq:ClDuStrin}
\end{equation}
where we have neglected a constant factor. The simplest (yet non-trivial) way to satisfy it is to assume that there is a symmetric positive definite fourth-rank \emph{dissipation tensor} $\bbD$, such that the evolution of the natural configuration is governed by the following gradient-flow equation 
\begin{equation}
\bbD (\bBeC) + \bPsi^{-1} - \bBe^{-1} = 0.
\label{eq:Diss}
\end{equation}

\section{Identification of the relaxation times}
\label{sec:rel-times}
It is natural to assume that the dissipation tensor must share the uniaxial symmetry of the shape tensor. The most general matrix representation of a symmetric fourth-rank tensor that is transversely isotropic about $\bn$ has five independent parameters \cite{87povi}. We interpret $\bbD:\Sym \to \Sym$ as an automorphism of the space of symmetric double tensors, $\Sym$. It is then natural to choose the following basis of $\Sym$
\begin{subequations}
\begin{align}
\bE_1 & = \frac{1}{\sqrt{2}}(\be_2 \tp \bn + \bn \tp \be_2) , \\
\bE_2 & = \frac{1}{\sqrt{2}}(\be_1 \tp \bn + \bn \tp \be_1) \\
\bE_3 & = \frac{1}{\sqrt{2}}(\be_1 \tp \be_2 + \be_2 \tp \be_1), \\
\bE_4 & = \frac{1}{\sqrt{2}}(\be_1 \tp \be_1 - \be_2 \tp \be_2) \\
\bE_5 & = \sqrt{\frac{3}{2}}\Big(\bn \tp \bn - \frac{1}{3} \bI\Big), \quad
\bE_6 = \frac{1}{\sqrt{3}}\bI .
\end{align}
\label{eq:baseE}
\end{subequations}
which is adapted from the orthonormal standard basis ($\be_1$,$\be_2$,$\bn$) of $\R^3$, where the $z$-axis is chosen along the director: $\be_3\equiv \bn$. The basis $\{\bE_i\}$ is orthonormal with respect to the dot product between double tensors, defined as 
\begin{equation}
\bL \cdot \bM = \tr(\bM^{\top}\bL),
\label{eq:scalar1}
\end{equation}
for any $\bL$, $\bM$ in $\Sym$. 

It will be apparent, when we study the propagation of an acoustic wave, that the relaxation times are not directly related to $\bbD$ but rather to the eigenvalues of the fourth-rank tensor $\bbT$, defined as 
\begin{equation}
\bbT = (\bPsi \tp \bPsi)\, \bbD, 
\label{eq:relaxation_tensor}
\end{equation}
where the tensor (or Kronecker) product between two double tensors is a fourth-rank tensor such that  
\begin{equation}
(\bL \tp \bM)\,\bX = \bL \bX \bM^{T} \, \text{ for all } \bL, \bM, \bX,
\end{equation}
and then we extend its definition by linearity. The tensor $\bbT$ should be compatible with the uniaxial symmetry about $\bn$, but it is not necessary symmetric with respect to the scalar product \eqref{eq:scalar1}. On the contrary, it must have a non-vanishing skew-symmetric part to provide the symmetry of $\bbD$. However, $\bbT$ is \emph{symmetric} with respect to the scalar product $\langle \cdot, \cdot\rangle$ defined as
\begin{equation}
\langle \bL ,\bM \rangle = \bL \cdot (\bPsi^{-1} \tp \bPsi^{-1})\bM = \bL \cdot \bPsi^{-1}\bM\bPsi^{-1}, 
\label{eq:scalar2}
\end{equation}
for any $\bL$, $\bM$ in $\Sym$ \footnote{The scalar product $\left< \cdot , \cdot\right>$ is well defined when $a(\rho)>0$ since the tensor $\bPsi^{-1} \tp \bPsi^{-1}$ is then positive definite and symmetric}. In fact, it is easy to check that $\langle \bL ,\bbT \bM \rangle = \langle \bbT\bL ,\bM \rangle$. Therefore, we can apply the spectral theorem to diagonalize $\bbT$. The eigenvectors of $\bbT$ are orthogonal with respect to $\langle \cdot, \cdot\rangle$. To further illustrate this point, let us introduce a $\langle \cdot, \cdot\rangle$-orthonormal basis, derived from $\{\bE_i\}$, which explicitly embeds the shape tensor $\bPsi$
\begin{subequations}
\begin{align}
\bL_1 & = \sqrt{a(\rho)}\bE_1, \quad \bL_2 = \sqrt{a(\rho)}\bE_2, \\
\bL_3 & = a(\rho)^{-1}\bE_3, \quad \bL_4 = a(\rho)^{-1}\bE_4, \\
\bL_5 & = \sqrt{\frac{2}{3}}\Big(a(\rho)^2\bn \tp \bn - \frac{1}{2a(\rho)} (\bI-\bn \tp \bn)\Big), \\
\bL_6 & = \frac{1}{\sqrt{3}}\bPsi .
\end{align}
\label{eq:baseL}
\end{subequations}
The most general relaxation tensor $\bbT$ that shares the symmetry of the shape tensor $\bPsi$ may be parameterized by five scalar coefficients $\tau_1$, $\tau_2$, $\tau_3$, $\tau_4$ and $\ang$. Its matrix representation using the basis $\{\bL_i\}$ is
\begin{align*}[\bbT]_{\bL} = 
\begin{pmatrix}
\tau_1 & 0 & 0 & 0 & 0 & 0 \\
0 & \tau_1 & 0 & 0 & 0 & 0 \\
0 & 0 & \tau_2 & 0 & 0 & 0 \\
0 & 0 & 0 & \tau_2 & 0 & 0 \\
0 & 0 & 0 & 0 & \taus+\taud\cos(2\ang) & \taud\sin(2\ang) \\
0 & 0 & 0 & 0 & \taud\sin(2\ang) & \taus-\taud\cos(2\ang)
\end{pmatrix},
\end{align*}
where for compactness we have introduced the notation
\begin{equation}
\taus = \frac{1}{2}(\tau_3 + \tau_4), \qquad  \taud = \frac{1}{2}(\tau_3 - \tau_4).
\end{equation}
The relaxation times $\tau_1$, $\tau_2$, $\tau_3$ and $\tau_4$ are the eigenvalues of $\bbT$, while $\ang$ is the angle between the eigenspace associated to $\tau_3$ and $\bL_5$. If $\ang=0$, $[\bbT]_{\bL}$ is diagonal. However, in general, $\ang$ is an additional parameter of the model. 

The spectral decomposition of $\bbT$ is 
\begin{equation}
\bbT = \sum_{p,q=1}^{6} [T_{pq}]_{\bL} \,\bL_p \,\kpp\, \bL_q, 
\label{eq:spectral}
\end{equation}
where the outer product $\kpp$ is such that the following identity holds
\begin{equation}
(\bL \,\kpp\, \bM)\,\bX = \bL\, \langle \bM , \bX \rangle, 
\label{eq:outer_star}
\end{equation}
for any $\bL$, $\bM$ and $\bX$ in $\Sym$. The matrix entries of $\bbT$ with respect to $\{\bE_i\}$ are derived from Eqs.\eqref{eq:spectral},\eqref{eq:outer_star} and Eqs.\eqref{eq:baseE},\eqref{eq:baseL}
\begin{equation}
[T_{ij}]_{\bE} = \bE_i \cdot \bbT \bE_j
= \sum_{p,q}(\bE_i \cdot \bL_p) [T_{pq}]_{\bL} \langle \bL_q , \bE_j \rangle .
\end{equation}
After some algebraic simplifications, which we omit for brevity, we finally obtain the expression of $\bbT$ that is convenient to use during the calculations
\begin{widetext}
\begin{align}[\bbT]_{\bE} = 
\begin{pmatrix}
\tau_1 & 0 & 0 & 0 & 0 & 0 \\
0 & \tau_1 & 0 & 0 & 0 & 0 \\
0 & 0 & \tau_2 & 0 & 0 & 0 \\
0 & 0 & 0 & \tau_2 & 0 & 0 \\
0 & 0 & 0 & 0 & \taus + \frac{\taud \left(\left(4 a^6+a^3+4\right) \cos (2 \ang )-\sqrt{2} \left(a^3-1\right)^2 \sin (2 \ang )\right)}{9 a^3}
&
\frac{\taud \left(\left(2 a^6+8 a^3-1\right) \sin (2 \ang )+2 \sqrt{2} \left(-2 a^6+a^3+1\right) \cos (2 \ang )\right)}{9 a^3}
\\
0 & 0 & 0 & 0 & \frac{\taud \left(\left(-a^6+8 a^3+2\right) \sin (2 \ang )+2 \sqrt{2} \left(a^6+a^3-2\right) \cos (2 \ang )\right)}{9 a^3}
&
\taus + \frac{\taud \left(\sqrt{2} \left(a^3-1\right)^2 \sin (2 \ang )-\left(4 a^6+a^3+4\right) \cos (2 \ang )\right)}{9 a^3}
\end{pmatrix} \notag
\end{align}
\end{widetext}
Eq.\eqref{eq:relaxation_tensor} can then be inverted to yield the dissipation tensor
\begin{equation}
\bbD = (\bPsi^{-1} \tp \bPsi^{-1})\bbT.
\end{equation}
The mathematical structure described above shows that $\bbD$ is positive definite if and only if $\bbT$ is positive definite, i.e, if all the relaxation times are strictly greater than zero
\begin{equation}
\tau_1 >0, \qquad \tau_2 >0, \qquad \tau_3>0, \qquad \tau_4>0,
\end{equation}
a condition that it is very natural to take for granted.

\section{Weak-flow approximation}
\label{sec:weak-flow}
When the relaxation mechanisms are much faster then the macroscopic dynamics, the material effectively behaves as a fluid and the model is expected to reduce to the compressible Ericksen-Leslie theory. An analysis of this approximation in the incompressible case has been presented in \cite{16bdt}. Here, we revise it to include compressibility. 

The characteristic time of deformation is related to the velocity gradient, so that we posit $\taudef=1/\|\nabla \bv\|$. This has to be compared with the overall relaxation time, which we can define as $\taur = \|\bbT\|$. We want to study the asymptotic approximation of the model in the limit $\taur \ll \taudef$, i.e.,
\[\taur \|\nabla \bv\| = \eps \ll 1, \]
where $\eps$ is a small parameter. This is a ``weak-flow approximation'' where the material reorganization is much faster than the deformation. Therefore, the effective tensor $\bBe$ differs from its equilibrium value only by a small amount
\begin{equation}
\bBe = \bPsi + \bB_1,\qquad \text{ with } \quad \|\bB_1\| = O(\eps).
\label{eq:approx_lin}
\end{equation}
By inserting \eqref{eq:approx_lin} into the evolution equation \eqref{eq:Diss}, to first order we find
\begin{equation}
\bPsi^{-1}\bB_1 = - \bbD (\bPsiC)\bPsi,
\end{equation}
which yields the following approximation for the stress tensor
\begin{equation}
\bT = -p\bI - \rho\mu(\rho) \bbD (\bPsiC)\bPsi, 
\label{eq:Tstress2} 
\end{equation}
with
\begin{equation}
p = \rho^2 \D{\sio}{\rho} 
+ \rho^2\mu(\rho ) \frac{3 a'(\rho)}{2a(\rho)} \dev(\bn\tp\bn) \cdot (\bbD (\bPsiC)\bPsi) .
\end{equation}
The co-deformational derivative of the shape tensor is calculated as
\begin{equation}
\begin{split}
\bPsiC & = \frac{\rho a'(\rho)}{a(\rho)^2} (\tr\bD) \left(\bI - (1+2a(\rho)^3)(\bn\tp\bn) \right) \\
& + \big(a(\rho)^2 - a(\rho)^{-1} \big)\big(\bno\tp\bn + \bn \tp \bno \\
& - \bD\bn\tp\bn- \bn\tp\bD\bn \big) - 2a(\rho)^{-1}\,\bD,
\end{split}
\label{eq:bLup}
\end{equation}
where $\bno=\dot{\bn}-\bW\bn$ is the co-rotational derivative of the nematic director, $\bD=\sym(\bna\bv)$ is the stretching, and $\bW=\skw(\bna\bv)$ is the spin.
We are now in a position to compare our result \eqref{eq:Tstress2} with the most general linear viscous stress compatible with the nematic structure, as given in the compressible Ericksen-Leslie theory, namely
\begin{equation}
\begin{split}
\bT_{\text{EL}} & = -\piso\bI + \alpha_1 (\bn\cdot \bD\bn)(\bn\tp\bn) + \alpha_2 (\bno \tp  \bn) \\
& + \alpha_3 (\bn \tp  \bno) + \alpha_4 \bD + \alpha_5 (\bD\bn \tp \bn) + \alpha_6 (\bn \tp \bD\bn) \\
& + \alpha_7 \big((\tr \bD ) (\bn\tp\bn) + (\bn\cdot \bD\bn)\bI \big) + \alpha_8 (\tr \bD ) \bI,
\end{split}
\label{eq:TLeslie}
\end{equation}
where $\piso(\rho) = \rho^2\sio'(\rho)$ is the isotropic pressure function. After some algebra, the Leslie coefficients are identified as follows
\begin{subequations}
\begin{align}
\alpha_1 & = \rho\mu \Big(\tau_2 - \frac{\left(a^3+1\right)^2}{a^3}\tau_1 + 3 \tau_3 (\cos\ang)^2 \notag\\
& + 3\tau_4 (\sin\ang)^2\Big), \\
\alpha_2 & = - \rho\mu \left(a^3-1\right) \tau_1, \\
\alpha_3 & =  - \rho\mu \left(1 - a^{-3}\right) \tau_1 ,  \\
\alpha_4 & = 2 \rho\mu \tau_2,  \\
\alpha_5 & = \rho \mu \Big(\left(1+a^3\right)\tau_1 - 2 \tau_2 \Big), \\
\alpha_6 & = \rho \mu  \Big(\left(1+a^{-3}\right)\tau_1 - 2 \tau_2 \Big), 
\end{align}
\begin{align}
\alpha_7 & = \rho \mu \Big(\tau_2 
+ \tau_3 \cos\ang \left((3 \kappa -1) \cos (\ang )+\sqrt{2} \sin \ang \right) \notag \\
& + \tau_4 \sin\ang \left((3 \kappa -1) \sin (\ang )-\sqrt{2} \cos \ang \right)
\Big) , \\
\alpha_8 & = \rho\mu \Big(- \tau_2 + \frac{1}{6} \tau_3 \big((9 \kappa ^2-6 \kappa -1) \cos (2 \ang ) \notag \\
& +9 \kappa ^2+2 \sqrt{2} (3 \kappa -1) \sin (2 \ang )-6 \kappa +3\big) \notag \\
& - \frac{1}{6} \tau_4 \big((9 \kappa ^2-6 \kappa -1) \cos (2 \ang ) \notag \\
& -9 \kappa ^2+2 \sqrt{2} (3 \kappa -1) \sin (2 \ang )+6 \kappa -3\big) \Big),
\end{align}
\label{eq:alphas2}
\end{subequations}
where we have defined $\kappa(\rho) = \rho  a'(\rho)/a(\rho)$. These viscosity coefficients satisfy identically the well-known Parodi relation \cite{70paro}
\begin{equation}\label{eq:P}
\alpha_2+\alpha_3=\alpha_6-\alpha_5\,,
\end{equation}
and the less obvious identities \cite{16bdt}
\begin{equation}\label{eq:P-like}
\frac{\alpha_2}{\alpha_3}=\frac{\alpha_4+\alpha_5}{\alpha_4+\alpha_6}=a^3.
\end{equation}

\section{Acoustic approximation}
\label{sec:acoustic}
In Sec.\ref{sec:weak-flow} we have studied the approximation of our model for weak flows and large deformations. We now analyze the ``dual'' approximation, where the deformations are small and the NLC is only weakly compressible, but the shear rates are virtually large. In Ref.\cite{14bdt,15tur} we have studied a similar problem, under the simplifying assumption of a single relaxation time and nearly-isotropic shape tensor, i.e., $a(\rho) \approx 1$. Here, we reconsider the nemato\-acoustic problem we have tackled in \cite{14bdt} but in the more general framework of Sec.\ref{sec:dynamics}, with the four relaxation times allowed by symmetry and arbitrary large anisotropies of the shape tensor.

A plane wave solution is represented by a displacement field of the form
\begin{subequations}
\begin{align}
\bu(\bx,t) & = \eps \ba \re\big\{\er^{i\phi(\bx,t)} \big\}, \\
\phi(\bx,t) & = \bkc \cdot \bx-\omega t , \qquad \bkc = \bk + i \bl,
\end{align}
\label{eq:ansatz_acoustic}
\end{subequations}
where $\phi(\bx,t)$ is a complex phase, $\bkc$ is a complex wave vector, $\omega$ is the angular frequency, $\eps\ll 1$ is a dimensionless small parameter that scales the amplitude of the wave, and the vector $\ba$ determines the amplitude and the polarization of the wave (e.g., longitudinal/transversal if $\ba$ is parallel/orthogonal to $\bk$). The real part of $\bkc$ is the ordinary wave vector $\bk=k \be$, with $\be$ a unit vector along the propagation direction. The imaginary part of $\bkc$ determines the attenuation of the wave. For later convenience in the calculations, we will retain the complex notation with the implicit understanding that only the real part of the equations has a physical meaning.

To order $O(\eps)$, the ansatz \eqref{eq:ansatz_acoustic} implies 
\begin{subequations}
\begin{align}
\bv & = -i \omega \eps \ba \er^{i\phi} , \qquad \dot{\bv} = - \omega^2 \eps \ba \er^{i\phi} \\
\bF & = \bI + \eps \bF_1 = \bI + i \eps \er^{i\phi}(\ba \tp \bkc), \\
\rho & = \rho_0 + \eps \rho_1 = \rho_0 \big(1 - i \eps \er^{i\phi}(\ba \cdot \bkc) \big), \label{eq:density_eps}\\
a(\rho) & = a_0 + \eps \frac{a_1}{\rho_0} \rho_1 , \quad a'(\rho_0) = a_1/\rho_0.
\end{align}
\label{eq:acoustic_eps1}
\end{subequations}
The density-dependent shape tensor and dissipation tensor are perturbed analogously. We omit their explicit expressions that can be easily calculated from Eqs.\eqref{eq:acoustic_eps1}, but we write formally $\bPsi = \bPsi_0 + \eps \bPsi_1$ and $\bbD = \bbD_0 + \eps \bbD_1$.

Our perturbative approach based on small density variations is particularly suited to analyze the quasi-incompressible response of liquids. In this approximation, tiny density variations should imply fairly large pressure changes. Specifically, the isotropic pressure function $\piso(\rho) = \rho^2\sio'(\rho)$ is expanded as
\[\piso(\rho_0 + \eps \rho_1) = p_0 + \eps p_1 \rho_1 , \]
with $p_0=\piso(\rho_0)$ and $\piso'(\rho_0)=p_1$. We recall that $v_0=\sqrt{p_1}$ is the (isotropic) sound speed in ordinary liquids, and $\rho_0 p_1$ is usually referred to as the bulk modulus. It measures the material response to compression. The pressure variations are $\Delta p \approx p_1 \Delta \rho$ and quasi-incompressibility implies that $\Delta p/p_0 \gg \Delta \rho/\rho_0$, i.e, $\rho_0 p_1 \gg p_0$. On the other hand, for the asymptotic procedure to be successful, it is required that $p_0 \gg \eps p_1 \rho_1$. Hence, we assume
\begin{equation}
\rho_0 p_1 \gg p_0 \gg \eps p_1 \rho_1 . 
\end{equation}
The evolution equation \eqref{eq:Diss} controls the relaxation mechanisms of the fluids. Its asymptotic analysis is best studied by means of the inverse relaxing strain $\bH = (\bG^{\top}\bG)^{-1}$, so that the effective strain can be written as $\bBe=\bF\bH\bF^{\top}$. To order $O(\eps)$, we find 
\begin{align}
\bH & = \bPsi_0 + \eps (\bH_1 + \bPsi_1) \label{eq:H}\\
\bBeC & = \bF\Ldot{\bH}\bF^T = \eps (\Ldot{\bH}_1 + \Ldot{\bPsi}_1)
\end{align}
where $\bH_1$ is an unknown tensor to be determined by solving Eq.\eqref{eq:Diss}. To first order, and after the transient has died out, this equation yields
\begin{equation}
\begin{split}
\big(\bbI &- i\omega(\bPsi_0\tp\bPsi_0)\bbD_0\big) \bH_1 \\
& = - \bF_1\bPsi_0  - \bPsi_0\bF_1^{T} + i\omega(\bPsi_0\tp\bPsi_0)\bbD_0(\bPsi_1), 
\end{split}
\label{eq:H_1}
\end{equation}
where $\bbI$ is the fourth-rank identity tensor. This equation clearly shows that the normal modes of relaxation are related to the eigenmodes of the tensor $\bbI - i\omega(\bPsi_0\tp\bPsi_0)\bbD_0$. The fourth-rank time relaxation tensor is then found to be $\bbT=(\bPsi\tp\bPsi)\bbD$, in agreement with Eq.\eqref{eq:relaxation_tensor}.

From Eqs.\eqref{eq:shape}-\eqref{eq:pressure}, and Eqs.\eqref{eq:H}-\eqref{eq:H_1} we readily obtain the stress tensor $\bT$. The calculations are straightforward but very lengthy, so they will not be reported here. They are best automated with a computer algebra software such as \textsc{mathematica\textsuperscript{tm}}. However, it is interesting to observe that by comparing the stress tensor $\bT$ with $\bT_{\text{EL}}$ as given in Eq.\eqref{eq:TLeslie}, we derive the frequency dependence of the Leslie viscosities predicted by our model. The low-frequency limit ($\omega=0$) coincides with the previous result \eqref{eq:alphas2}, which was obtained by quite a different procedure. We find
\begin{subequations}
\begin{align}
\alpha_1 & =  \rho_0\mu_0 \bigg(-\frac{(a_0^{3}+1)^2}{a_0^3}  \frac{\tau_1}{1+(\omega\tau_1)^2} + \frac{\tau_2}{1 + (\omega\tau_2)^2} \notag\\
& + C_{13}\frac{\tau_3}{1 + (\omega\tau_3)^2}
+ C_{14}\frac{\tau_4}{1 + (\omega\tau_4)^2}\bigg), \\
\alpha_2 & = -\rho_0\mu_0 (a_0^3-1)\,  \frac{\tau_1}{1+(\omega\tau_1)^2}, \\
\alpha_3 & = -\rho_0\mu_0 (1-a_0^{-3})\,  \frac{\tau_1}{1+(\omega\tau_1)^2}, \\
\alpha_4 & =  \rho_0\mu_0 \frac{2\tau_2}{1 + (\omega\tau_2)^2}, \\
\alpha_5 & = \rho_0\mu_0 \left(\frac{(a_0^3 + 1)\tau_1}{1+(\omega\tau_1)^2} - \frac{2\tau_2}{1 + (\omega\tau_2)^2}\right), \\
\alpha_6 & = \rho_0\mu_0 \left(\frac{(1+a_0^{-3})\tau_1}{1+(\omega\tau_1)^2} - \frac{2\tau_2}{1 + (\omega\tau_2)^2}\right), 
\end{align}
\begin{align}
\alpha_7 & = \rho_0\mu_0 \bigg(\frac{\tau_2}{1+(\omega\tau_2)^2} + C_{73}\frac{\tau_3}{1+(\omega\tau_3)^2} \notag\\
&+ C_{74}\frac{\tau_4}{1+(\omega\tau_4)^2}\bigg), \\
\alpha_8 & = \rho_0\mu_0 \bigg(-\frac{\tau_2}{1+(\omega\tau_2)^2} + C_{83}\frac{\tau_3}{1+(\omega\tau_3)^2} \notag \\
&+ C_{84}\frac{\tau_4}{1+(\omega\tau_4)^2}\bigg),
\end{align}
\label{eq:alpha_acustico}
\end{subequations}
where
\begin{subequations}
\begin{align}
C_{13} & = 3(\cos\ang)^2, \qquad C_{14} = 3(\sin\ang)^2, \\
C_{73} & = \cos\ang \left((3 \kappa_0-1) \cos\ang
+\sqrt{2} \sin \ang \right), \\
C_{74} & = \sin\ang \left((3 \kappa_0 -1)\sin\ang
-\sqrt{2} \cos\ang\right), \\
C_{83} & = \big(2 \sqrt{2}\left(3\kappa_0-1\right)\sin (2 \ang) \notag \\
& + \left(9 \kappa _0^2-6 \kappa_0-1\right) \cos (2 \ang ) \notag \\
& + 9 \kappa_0^2-6 \kappa _0+3\big)/6, \\
C_{84} & = \big(-2 \sqrt{2}\left(3\kappa_0-1\right)\sin(2\ang) \notag \\
& -\left(9 \kappa_0^2-6 \kappa_0-1\right)\cos (2 \ang ) \notag \\
& + 9 \kappa_0^2-6 \kappa_0+3\big)/6,
\end{align}
\end{subequations}
and $\mu_0 = \mu(\rho_0)$, $\kappa_0 = a_1/a_0$. The identities \eqref{eq:P} and \eqref{eq:P-like} still hold true at any frequency.

\subsection{Solution of the Christoffel equation}
In order to investigate the propagation of sound we have to solve the balance of momentum equation (\ref{eq:balances}b). In the acoustic approximation, to first order this reads
\begin{equation}
-\rho_0 \omega^2 \ba \er^{i\phi} = i \bT \,\bkc.
\label{eq:Christoffel1}
\end{equation}
The unknowns of this complex equation are: the amplitude vector $\ba$ that determines the polarization of the sound wave, the sound phase velocity $\vs=\omega/k$, and the attenuation vector $\bl$. However, the solution of Eq.\eqref{eq:Christoffel1} is obscured by the fact that the amplitude vector $\ba$ is implicitly contained in the stress tensor. To make the mathematical structure of Eq.\eqref{eq:Christoffel1} more transparent, we observe that $\bT$ depends linearly on $\ba$ so that we can define two double tensors $\bA_r$ and $\bA_i$ such that
\begin{equation}
- \frac{\er^{-i\phi}}{\rho_0} i\bT\,\bkc = (\bA_r+i\bA_i)\ba.
\end{equation}
Eq.\eqref{eq:Christoffel1} can now be recast in the form of a complex eigenvalue problem, whose real and imaginary parts are
\begin{equation}
\begin{cases}
\bA_r \ba = \omega^2\, \ba, \\
\bA_i \ba = 0,
\end{cases}
\label{eq:Christoffel2}
\end{equation}
Eq.\eqref{eq:Christoffel2} is known as the Christoffel equation in the theory of acoustic propagation in solids \cite{68Fedorov}. Its real part, Eq.(\ref{eq:Christoffel2}a), determines the direction of polarization of the wave and the sound speed. The imaginary part, Eq.(\ref{eq:Christoffel2}b), identifies the attenuation.

The Christoffel equation is still too complicated to be solved directly. One simplifying assumption, very natural in our context, is that the stored energy density \eqref{eq:StrainEnergy} is only weakly anisotropic and it is dominated by its first term \cite{14bdt}. In mathematical terms, we posit that the shear modulus $\rho_0\mu_0$ is much smaller than the bulk modulus $\rho_0 p_1$, so that their ratio is small
\begin{equation}
\eta = \mu_0/p_1 \ll 1.
\label{eq:WeakAnisotropy}
\end{equation}

\subsubsection{Sound speed}
According to Eq.(\ref{eq:Christoffel2}a), the amplitude vector $\ba$ must be an eigenvector of $\bA_r$. Since the matrix $\bA_r$ depends on the wave number $k$, we can determine the sound speed $\vs=\omega/k$ by imposing that $\omega^2$ is an eigenvalue of $\bA_r$. Hence, the equation for the sound speed is 
\begin{equation}
\det(\bA_r - \omega^2 \bI) = 0,
\label{eq:vs_1} 
\end{equation}
which can be solved without a prior knowledge of the polarization $\ba$. Given the weak anisotropy assumption \eqref{eq:WeakAnisotropy}, we look for a solution of Eq.\eqref{eq:vs_1} in the form of an asymptotic expansion $\vs = v_0 + \eta v_1 + o(\eta)$, where $v_0 = \sqrt{p_1}$. After some algebra, we find
\begin{equation}
\begin{split}
\vs = v_0 & + \eta v_0 \Bigg[A^{(0)}_{0} + \sum_{i=1}^{4}\frac{A^{(0)}_{i}}{1+(\omega\tau_i)^2} \\
& + \left(A^{(2)}_{0} + \sum_{i=1}^{4}\frac{A^{(2)}_{i}}{1+(\omega\tau_i)^2} \right)\cos(2\theta)  \\
& + \left(A^{(4)}_{0} + \sum_{i=1}^{4}\frac{A^{(4)}_{i}}{1+(\omega\tau_i)^2} \right)\cos(4\theta) \Bigg],
\end{split}
\label{eq:soundspeed}
\end{equation}
with
\begin{subequations}
\begin{align}
A^{(0)}_{0} & = \left(a_0^3+a_0^{-3}+8 \kappa_0 (3 \kappa_0+1)+14\right)/16, \\
A^{(0)}_{1} & = -\left(a_0^3+1\right)^2/(16 a_0^3), \\
A^{(0)}_{2} & = -3/16,\\
A^{(0)}_{3} & = \Big(-8 \sqrt{2} \left(6\kappa_0+1\right) \sin (2 \ang) \notag \\
& + \left(5-24 \kappa _0 \left(3\kappa_0+1\right)\right) \cos (2\ang)\notag \\
& - 3 \left(8 \kappa_0 \left(3 \kappa_0+1\right) + 9 \right)\Big)/96, \\
A^{(0)}_{4} & = \Big(8 \sqrt{2} \left(6\kappa_0+1\right) 
\sin (2 \ang) \notag \\
& - \left(5-24 \kappa_0 \left(3\kappa_0+1\right)\right) \cos (2 \ang)\notag \\
& - 3 \left(8 \kappa_0 \left(3\kappa_0+1\right) + 9 \right)\Big)/96,
\end{align}
\begin{align}
A^{(2)}_{0} & = 3\kappa_0/2, \qquad A^{(2)}_{1} = 0, \qquad A^{(2)}_{2} = 1/4, \\
A^{(2)}_{3} & = -\cos\ang \big( (6\kappa_0+1)
\cos\ang + 2 \sqrt{2} \sin\ang \big)/4, \\
A^{(2)}_{4} & = -\sin\ang \big( (6\kappa_0+1) \sin\ang - 2 \sqrt{2} \cos\ang \big)/4,
\end{align}
\begin{align}
A^{(4)}_{0} & = -\left(a_0^3-1\right)^2/(16 a_0^3), \\
A^{(4)}_{1} & = \left(a_0^3+1\right)^2/(16 a_0^3), \\
A^{(4)}_{2} & = -1/16,\\
A^{(4)}_{3} & = -3(\cos\ang)^2/16, \qquad 
A^{(4)}_{4}   = -3(\sin\ang)^2/16.
\end{align}
\end{subequations}

The amplitude vector $\ba$ is the eigenvector of $\bA_r$ relative to the eigenvalue $\omega^2$. It is worth noticing that the matrix $\bA_r$ depends on the wave number $k=\omega/\vs$, with $\vs$ as given in Eq.\eqref{eq:soundspeed}. To leading order, we obtain, as expected, purely longitudinal waves, i.e., $\ba =A_0 \be$. The small $O(\eta)$-correction to the polarization is orthogonal to $\be$ and can be calculated explicitly, but its long awkward expression is not particularly illuminating and it will not be reported here for brevity.

\subsubsection{Sound attenuation}
Finally, the attenuation vector is found by solving Eq.(\ref{eq:Christoffel2}b), where $\bl$ enters implicitly into the definition of $\bA_i$. Since the wave is longitudinal to leading order, the attenuation vector $\bl$ is such that $\bA_i \be = 0$. This equation yields
\begin{align}
\frac{\bl}{\eta k_0} & = \sum_{i=1}^{4}\frac{\omega\tau_i}{1+(\omega\tau_i)^2}
\Big(B^{(0)}_{i}\!+\!B^{(2)}_{i}\cos(2\theta)\!+\!B^{(4)}_{i}\cos(4\theta)\Big)\be \notag \\
& + \sum_{i=1}^{4}\frac{\omega\tau_i}{1+(\omega\tau_i)^2} 
\Big(C^{(2)}_{i} \sin(2\theta) + C^{(4)}_{i} \sin(4\theta) \Big)\, \bt ,
\end{align}
where $k_0 = \omega/v_0$, and $\bt$ is the unit vector orthogonal to $\be$ and that belongs to the plane $\text{span}\{\be,\bn\}$, such that $\bn = \cos\theta \,\be + \sin\theta \,\bt$.  The coefficients are 
\begin{equation}
B^{(h)}_i = - A^{(h)}_i, \quad h=0,2,4;\,\, i=1,2,3,4.
\end{equation}

\begin{subequations}
\begin{align}
C^{(2)}_{1} & = -\left(a_0^6-1\right)/(4 a_0^3), \\
C^{(2)}_{i} & = -A^{(2)}_{i}, \quad i=2,3,4,
\end{align}
\begin{equation}
C^{(4)}_i = - 2A^{(4)}_i, \quad i=1,2,3,4.
\end{equation}
\end{subequations}

\section{Discussion}
We have developed a viscoelastic theory of nematic liquid crystals that accounts for both elastic and relaxation effects, based only on material symmetry requirements and compatibility with thermodynamics. The uniaxial symmetry implies that there are only four possible relaxation times. Using a weak-flow approximation, we are able to predict how the eight viscosities of the compressible Ericksen-Leslie model depend on these relaxation times. The acoustic approximation of our theory provides a number of further interesting results. First, we have obtained the frequency dependence of the eight viscosity coefficients, the sound speed and the attenuation. We have seen that each of these quantities is a superposition of relaxing terms, each of which is associated to a single relaxation time. In particular, according to Eqs.\eqref{eq:alphas2}, we have modeled NLCs as \emph{shear-thinning} fluids: at low shear rates, the shear stress is proportional to $\bna\bv$, and the viscosity approaches a constant value. At 
higher shear rates the viscosity decreases with increasing shear rate. This behavior is observed in most (but not all) polymeric liquids \cite{87bird}. Secondly, we have found how the sound speed and the acoustic attenuation depend on the angle between the director and the direction of propagation.

In principle, our results are in good qualitative agreement with experiments. In order to make this comparison quantitatively precise, it is desirable to fit the experimental measures and thus find the parameters of our model. However, at the present day, this seems to be an intractable task and only a rough estimate of the parameters can be aimed at. There are three main reasons for this. 

The first reason is theoretical: our theory does not account for partial order and temperature effects. By contrast, relaxation times, viscosity coefficients and acoustic properties are known to depend on temperature and degree of order. We plan to extend our theory to include the effects of partial order in a future paper. For the present purposes, we try to select the experimental data sharing the same degree of order and we obtain a crude reconstruction of the nominal values of nematic viscosities at perfect order $S = 1$ from values measured for partially ordered NLCs by replacing each $\bn \tp \bn$ term in the Cauchy stress tensor by the corresponding second-moment $S(\bn \tp \bn)$.

\vspace{0.2cm}
\begin{figure}[t]
\centering
\includegraphics[width=0.4\textwidth]{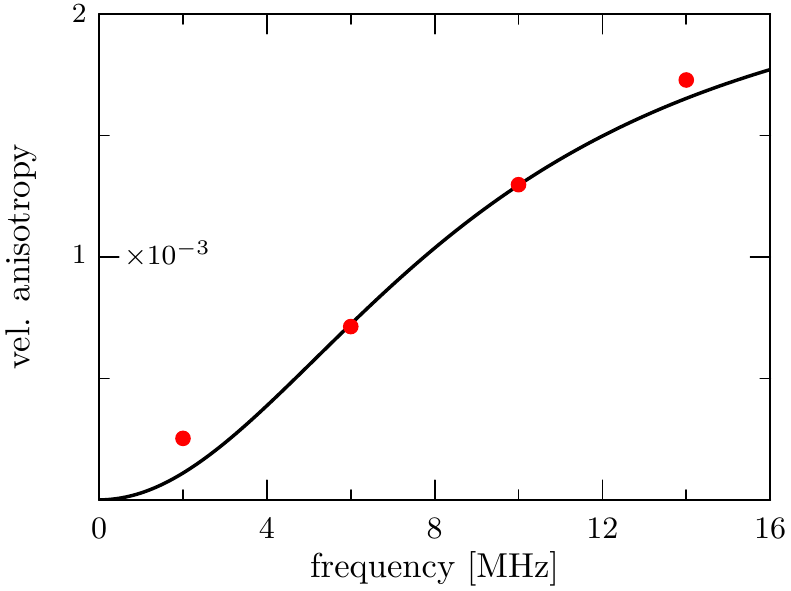}
\caption{Frequency dependence of the velocity anisotropy. The solid line represents our fit with the four relaxation times and the model parameters given in Tab.~\ref{tab:MBBA}. Experimental points are taken from \cite{72muls}.}
\label{fig:velocity}
\end{figure}

Secondly, the experimental data at our disposal are scattered in the literature among various sources from the early 70's. Therefore, usually the experimental settings that we have to cope with, may differ with respect to temperature, sample purity, material degradation, and accuracy of the experimental techniques.

Finally, ultrasonic measurements are usually performed on a limited range of frequencies, typically below $100$MHz. However, to be able to determine the relaxation times associated with the fast relaxation modes ($< 10^{-9}$s), we need to study the propagation of sound in the GHz-range. Modern techniques based on Brillouin scattering seem to enable the analysis of sound propagation up to tens of GHz. However, we were not able to find such measurements for MBBA in the right temperature range.

Consistently to what is done in Refs.\cite{14bdt,16bdt,09virga,11DeMatteis}, we will use what seems to be a reasonably coherent set of data for MBBA: (i) the shear viscosities as suggested in Ref.\cite{82kneppe} (except for $\alpha_3$, see \cite{16bdt}), at $T=25^\circ$C, corresponding to a degree of order $S=0.66$, according to \cite{16bdt}; (ii) the anisotropic profile of the sound speed at $2$, $6$, $10$, $14$MHz and $T=25^\circ$C as given in Fig.3 of Ref.\cite{72muls}; (iii) the attenuation anisotropy as given in Fig.2 of Ref.\cite{70lord}.


There are eight unknown parameters specific to our model: the shear modulus $\rho_0\mu_0$ (or, equivalently, $\eta$), the aspect ratio $a_0$, its compressibility $\kappa_0=a_1/a_0$, four relaxation times $\tau_1$, $\tau_2$, $\tau_3$, $\tau_4$, and the angle $\ang$. In order to simplify the fitting procedure, we have to reduce the dimension of the parameter space. To this end, we evaluate the aspect ratio $a_0$ by means of Eq.\eqref{eq:P-like}, using the experimental measure\cite{82kneppe} of $\alpha_4$ and the extrapolated values of $\alpha_5$, $\alpha_6$ from $S=0.66$ to $S=1$ (see also \cite{16bdt}). The values of $\alpha_4$ and $\alpha_5$ are also used to determine the products of the shear modulus, $\rho_0 \mu_0$, with $\tau_1$ and $\tau_2$. A further equation linking $\tau_3$, $\tau_4$ and $\ang$ is obtained from the $S^2$-rescaled value of $\alpha_1$.

We are now able to determine the remaining parameters, namely $\rho_0\mu_0$, $\kappa_0$, $\tau_4$ and $\ang$, by fitting the functional dependence of the velocity on the angle $\theta$ at four different frequencies \cite{72muls}. The best estimated model parameters are reported in Table \ref{tab:MBBA}. It is clear from Fig.\ref{fig:velocity} that the fit describes the velocity data quite well. The velocity anisotropy is defined as $\Delta v/v = \big(\vs(\theta=0)-\vs(\theta=\pi/2)\big)/\vs(\theta=\pi/2)$.

\vspace{0.2cm}
\begin{table}[h]
\centering
\begin{tabular}{|c|c|c|c|c|c|c|c|}\hline
$a_0$ & $\eta$ & $\kappa_0$ & $\tau_1$[ns] & $\tau_2$[ns] & $\tau_3$[ns] & $\tau_4$[ns] & $\ang$ \\[1mm] \hline
1.85 & 1.22 $\times 10^{-3}$ & 0.628 & 9.42 & 14.2 & 17.1 & 32.9 & 0.0 \\ \hline
\end{tabular}
\caption{Values of the model parameters identified by the experimental MBBA viscosity coefficients\cite{82kneppe} and a best fit of velocity anisotropy\cite{72muls}. See the text for details. The temperature of the samples is in all cases $25^\circ$C, corresponding to a degree of order $S\approx 0.66$. Standard values for the density ($\rho_0=10^3$kg/m$^3$) and the isotropic sound speed ($v_0=1540$m/s) are used.}
\label{tab:MBBA}
\end{table}

\vspace{0.2cm}
\begin{figure}[t]
\centering
\includegraphics[width=0.4\textwidth]{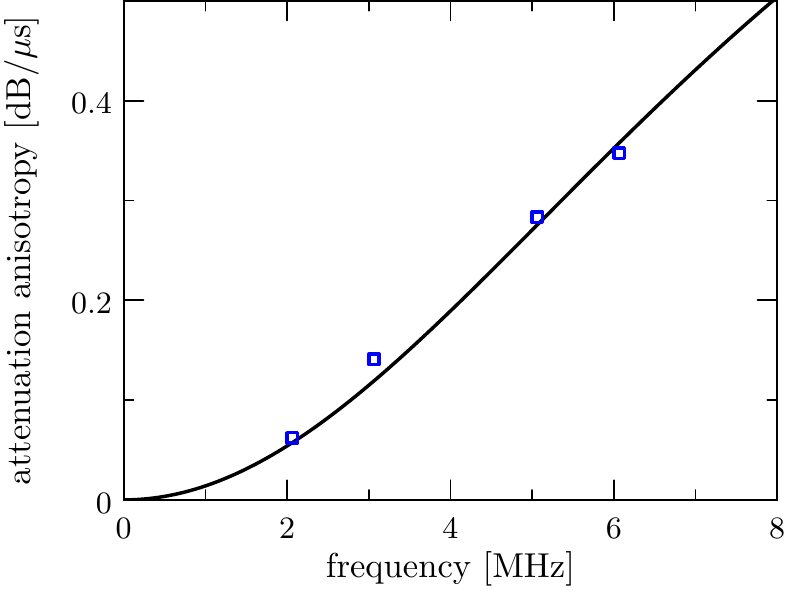}
\caption{Frequency dependence of the attenuation anisotropy. The solid line represents our theoretical estimate obtained with the model parameters reported in Tab.~\ref{tab:MBBA}. Experimental points are taken from \cite{70lord}.}
\label{fig:attenuation}
\end{figure}

Despite the unrefined data at hand and the necessary simplifying assumptions in the fitting procedure, the values in Tab.~\ref{tab:MBBA} have some predictive power. For example, $\alpha_2$ (which was not part of the optimization procedure) is found to be $-0.096$ Pa$\,$s, against an experimental value of $-0.11$ Pa$\,$s. The values of the bulk viscosities, $\alpha_7$ and $\alpha_8$, are predicted to be $0.056$ and $0.035$ Pa$\,$s respectively. Furthermore, we compare the theoretical attenuation anisotropy (in decibels per unit flight time) with the measured data of Ref.~\cite{70lord}, which were not included in the fit. The remarkable agreement of theory and experiment in this case is clearly displayed in Fig.~\ref{fig:attenuation}.

Finally, it is worth noticing that our theory should be able to cover with continuity a broad spectrum of possible behaviors, from isotropic viscous fluids to nematic elastomers. In particular, it should equally well be applicable to polymeric anisotropic fluids, i.e., macromolecular liquid crystalline polymers.

\begin{acknowledgments} It is a pleasure to thank Antonio Di Carlo and Paolo Biscari for many stimulating and helpful discussions.

\end{acknowledgments}


\begin{thebibliography}{38}%
\makeatletter
\providecommand \@ifxundefined [1]{%
 \@ifx{#1\undefined}
}%
\providecommand \@ifnum [1]{%
 \ifnum #1\expandafter \@firstoftwo
 \else \expandafter \@secondoftwo
 \fi
}%
\providecommand \@ifx [1]{%
 \ifx #1\expandafter \@firstoftwo
 \else \expandafter \@secondoftwo
 \fi
}%
\providecommand \natexlab [1]{#1}%
\providecommand \enquote  [1]{``#1''}%
\providecommand \bibnamefont  [1]{#1}%
\providecommand \bibfnamefont [1]{#1}%
\providecommand \citenamefont [1]{#1}%
\providecommand \href@noop [0]{\@secondoftwo}%
\providecommand \href [0]{\begingroup \@sanitize@url \@href}%
\providecommand \@href[1]{\@@startlink{#1}\@@href}%
\providecommand \@@href[1]{\endgroup#1\@@endlink}%
\providecommand \@sanitize@url [0]{\catcode `\\12\catcode `\$12\catcode
  `\&12\catcode `\#12\catcode `\^12\catcode `\_12\catcode `\%12\relax}%
\providecommand \@@startlink[1]{}%
\providecommand \@@endlink[0]{}%
\providecommand \url  [0]{\begingroup\@sanitize@url \@url }%
\providecommand \@url [1]{\endgroup\@href {#1}{\urlprefix }}%
\providecommand \urlprefix  [0]{URL }%
\providecommand \Eprint [0]{\href }%
\providecommand \doibase [0]{http://dx.doi.org/}%
\providecommand \selectlanguage [0]{\@gobble}%
\providecommand \bibinfo  [0]{\@secondoftwo}%
\providecommand \bibfield  [0]{\@secondoftwo}%
\providecommand \translation [1]{[#1]}%
\providecommand \BibitemOpen [0]{}%
\providecommand \bibitemStop [0]{}%
\providecommand \bibitemNoStop [0]{.\EOS\space}%
\providecommand \EOS [0]{\spacefactor3000\relax}%
\providecommand \BibitemShut  [1]{\csname bibitem#1\endcsname}%
\let\auto@bib@innerbib\@empty
\bibitem [{\citenamefont {{de Gennes}}\ and\ \citenamefont
  {Prost}(1995)}]{95dgpr}%
  \BibitemOpen
  \bibfield  {author} {\bibinfo {author} {\bibfnamefont {P.}~\bibnamefont {{de
  Gennes}}}\ and\ \bibinfo {author} {\bibfnamefont {J.}~\bibnamefont {Prost}},\
  }\href@noop {} {\emph {\bibinfo {title} {The Physics of Liquid Crystals}}},\
  \bibinfo {edition} {2nd}\ ed.\ (\bibinfo  {publisher} {Oxford University
  Press},\ \bibinfo {address} {Oxford},\ \bibinfo {year} {1995})\BibitemShut
  {NoStop}%
\bibitem [{\citenamefont {Mullen}\ \emph {et~al.}(1972)\citenamefont {Mullen},
  \citenamefont {L{\"{u}}thi},\ and\ \citenamefont {Stephen}}]{72muls}%
  \BibitemOpen
  \bibfield  {author} {\bibinfo {author} {\bibfnamefont {M.~E.}\ \bibnamefont
  {Mullen}}, \bibinfo {author} {\bibfnamefont {B.}~\bibnamefont {L{\"{u}}thi}},
  \ and\ \bibinfo {author} {\bibfnamefont {M.~J.}\ \bibnamefont {Stephen}},\
  }\href@noop {} {\bibfield  {journal} {\bibinfo  {journal} {Phys. Rev. Lett.}\
  }\textbf {\bibinfo {volume} {28}},\ \bibinfo {pages} {799} (\bibinfo {year}
  {1972})}\BibitemShut {NoStop}%
\bibitem [{\citenamefont {Miyano}\ and\ \citenamefont
  {Ketterson}(1975)}]{75Miyano}%
  \BibitemOpen
  \bibfield  {author} {\bibinfo {author} {\bibfnamefont {K.}~\bibnamefont
  {Miyano}}\ and\ \bibinfo {author} {\bibfnamefont {J.~B.}\ \bibnamefont
  {Ketterson}},\ }\href {\doibase 10.1103/PhysRevA.12.615} {\bibfield
  {journal} {\bibinfo  {journal} {Phys. Rev. A}\ }\textbf {\bibinfo {volume}
  {12}},\ \bibinfo {pages} {615} (\bibinfo {year} {1975})}\BibitemShut
  {NoStop}%
\bibitem [{\citenamefont {Grammes}\ \emph {et~al.}(1995)\citenamefont
  {Grammes}, \citenamefont {Kr\"uger}, \citenamefont {Bohn}, \citenamefont
  {Baller}, \citenamefont {Fischer}, \citenamefont {Schorr}, \citenamefont
  {Rogez},\ and\ \citenamefont {Alnot}}]{95gr}%
  \BibitemOpen
  \bibfield  {author} {\bibinfo {author} {\bibfnamefont {C.}~\bibnamefont
  {Grammes}}, \bibinfo {author} {\bibfnamefont {J.~K.}\ \bibnamefont
  {Kr\"uger}}, \bibinfo {author} {\bibfnamefont {K.-P.}\ \bibnamefont {Bohn}},
  \bibinfo {author} {\bibfnamefont {J.}~\bibnamefont {Baller}}, \bibinfo
  {author} {\bibfnamefont {C.}~\bibnamefont {Fischer}}, \bibinfo {author}
  {\bibfnamefont {C.}~\bibnamefont {Schorr}}, \bibinfo {author} {\bibfnamefont
  {D.}~\bibnamefont {Rogez}}, \ and\ \bibinfo {author} {\bibfnamefont
  {P.}~\bibnamefont {Alnot}},\ }\href {\doibase 10.1103/PhysRevE.51.430}
  {\bibfield  {journal} {\bibinfo  {journal} {Phys. Rev. E}\ }\textbf {\bibinfo
  {volume} {51}},\ \bibinfo {pages} {430} (\bibinfo {year} {1995})}\BibitemShut
  {NoStop}%
\bibitem [{\citenamefont {Kr\"uger}\ \emph {et~al.}(1995)\citenamefont
  {Kr\"uger}, \citenamefont {Grammes}, \citenamefont {Jim\'enez}, \citenamefont
  {Schreiber}, \citenamefont {Bohn}, \citenamefont {Baller}, \citenamefont
  {Fischer}, \citenamefont {Rogez}, \citenamefont {Schorr},\ and\ \citenamefont
  {Alnot}}]{95kr}%
  \BibitemOpen
  \bibfield  {author} {\bibinfo {author} {\bibfnamefont {J.~K.}\ \bibnamefont
  {Kr\"uger}}, \bibinfo {author} {\bibfnamefont {C.}~\bibnamefont {Grammes}},
  \bibinfo {author} {\bibfnamefont {R.}~\bibnamefont {Jim\'enez}}, \bibinfo
  {author} {\bibfnamefont {J.}~\bibnamefont {Schreiber}}, \bibinfo {author}
  {\bibfnamefont {K.-P.}\ \bibnamefont {Bohn}}, \bibinfo {author}
  {\bibfnamefont {J.}~\bibnamefont {Baller}}, \bibinfo {author} {\bibfnamefont
  {C.}~\bibnamefont {Fischer}}, \bibinfo {author} {\bibfnamefont
  {D.}~\bibnamefont {Rogez}}, \bibinfo {author} {\bibfnamefont
  {C.}~\bibnamefont {Schorr}}, \ and\ \bibinfo {author} {\bibfnamefont
  {P.}~\bibnamefont {Alnot}},\ }\href {\doibase 10.1103/PhysRevE.51.2115}
  {\bibfield  {journal} {\bibinfo  {journal} {Phys. Rev. E}\ }\textbf {\bibinfo
  {volume} {51}},\ \bibinfo {pages} {2115} (\bibinfo {year}
  {1995})}\BibitemShut {NoStop}%
\bibitem [{\citenamefont {Kim}\ \emph {et~al.}(2012)\citenamefont {Kim},
  \citenamefont {Kim}, \citenamefont {Ko},\ and\ \citenamefont {Kim}}]{12kkkk}%
  \BibitemOpen
  \bibfield  {author} {\bibinfo {author} {\bibfnamefont {J.~H.}\ \bibnamefont
  {Kim}}, \bibinfo {author} {\bibfnamefont {T.~H.}\ \bibnamefont {Kim}},
  \bibinfo {author} {\bibfnamefont {J.-H.}\ \bibnamefont {Ko}}, \ and\ \bibinfo
  {author} {\bibfnamefont {J.-H.}\ \bibnamefont {Kim}},\ }\href {\doibase
  10.3938/jkps.61.862} {\bibfield  {journal} {\bibinfo  {journal} {J. Korean
  Phys. Soc.}\ }\textbf {\bibinfo {volume} {61}},\ \bibinfo {pages} {862}
  (\bibinfo {year} {2012})}\BibitemShut {NoStop}%
\bibitem [{\citenamefont {Wetsel~Jr.}\ \emph {et~al.}(1972)\citenamefont
  {Wetsel~Jr.}, \citenamefont {Speer}, \citenamefont {Lowry},\ and\
  \citenamefont {Woodard}}]{72wetsel}%
  \BibitemOpen
  \bibfield  {author} {\bibinfo {author} {\bibfnamefont {G.~C.}\ \bibnamefont
  {Wetsel~Jr.}}, \bibinfo {author} {\bibfnamefont {R.~S.}\ \bibnamefont
  {Speer}}, \bibinfo {author} {\bibfnamefont {B.~A.}\ \bibnamefont {Lowry}}, \
  and\ \bibinfo {author} {\bibfnamefont {M.~R.}\ \bibnamefont {Woodard}},\
  }\href {\doibase 10.1063/1.1661348} {\bibfield  {journal} {\bibinfo
  {journal} {J. Appl. Phys.}\ }\textbf {\bibinfo {volume} {43}},\ \bibinfo
  {pages} {1495} (\bibinfo {year} {1972})}\BibitemShut {NoStop}%
\bibitem [{\citenamefont {J{\"a}hnig}(1973)}]{73Jahnig}%
  \BibitemOpen
  \bibfield  {author} {\bibinfo {author} {\bibfnamefont {F.}~\bibnamefont
  {J{\"a}hnig}},\ }\href {\doibase 10.1007/BF01392830} {\bibfield  {journal}
  {\bibinfo  {journal} {Z. Physik}\ }\textbf {\bibinfo {volume} {258}},\
  \bibinfo {pages} {199} (\bibinfo {year} {1973})}\BibitemShut {NoStop}%
\bibitem [{\citenamefont {Eden}\ \emph {et~al.}(1973)\citenamefont {Eden},
  \citenamefont {Garland},\ and\ \citenamefont {Williamson}}]{73eden}%
  \BibitemOpen
  \bibfield  {author} {\bibinfo {author} {\bibfnamefont {D.}~\bibnamefont
  {Eden}}, \bibinfo {author} {\bibfnamefont {C.~W.}\ \bibnamefont {Garland}}, \
  and\ \bibinfo {author} {\bibfnamefont {R.~C.}\ \bibnamefont {Williamson}},\
  }\href@noop {} {\bibfield  {journal} {\bibinfo  {journal} {J. Chem. Phys.}\
  }\textbf {\bibinfo {volume} {58}},\ \bibinfo {pages} {1861} (\bibinfo {year}
  {1973})}\BibitemShut {NoStop}%
\bibitem [{\citenamefont {Kemp}\ and\ \citenamefont {Letcher}(1974)}]{74kemp}%
  \BibitemOpen
  \bibfield  {author} {\bibinfo {author} {\bibfnamefont {K.~A.}\ \bibnamefont
  {Kemp}}\ and\ \bibinfo {author} {\bibfnamefont {S.~V.}\ \bibnamefont
  {Letcher}},\ }\enquote {\bibinfo {title} {Bulk viscosities of mbba from
  ultrasonic measurements},}\ in\ \href {\doibase 10.1007/978-1-4684-2727-1_32}
  {\emph {\bibinfo {booktitle} {Liquid Crystals and Ordered Fluids: Volume
  2}}},\ \bibinfo {editor} {edited by\ \bibinfo {editor} {\bibfnamefont
  {J.~F.}\ \bibnamefont {Johnson}}\ and\ \bibinfo {editor} {\bibfnamefont
  {R.~S.}\ \bibnamefont {Porter}}}\ (\bibinfo  {publisher} {Springer US},\
  \bibinfo {address} {Boston, MA},\ \bibinfo {year} {1974})\ pp.\ \bibinfo
  {pages} {351--356}\BibitemShut {NoStop}%
\bibitem [{\citenamefont {Monroe~Jr.}\ \emph {et~al.}(1975)\citenamefont
  {Monroe~Jr.}, \citenamefont {Wetsel~Jr.}, \citenamefont {Woodard},\ and\
  \citenamefont {Lowry}}]{75monroe}%
  \BibitemOpen
  \bibfield  {author} {\bibinfo {author} {\bibfnamefont {S.~E.}\ \bibnamefont
  {Monroe~Jr.}}, \bibinfo {author} {\bibfnamefont {G.~C.}\ \bibnamefont
  {Wetsel~Jr.}}, \bibinfo {author} {\bibfnamefont {M.~R.}\ \bibnamefont
  {Woodard}}, \ and\ \bibinfo {author} {\bibfnamefont {B.~A.}\ \bibnamefont
  {Lowry}},\ }\href@noop {} {\bibfield  {journal} {\bibinfo  {journal}
  {J.~Chem.~Phys.}\ }\textbf {\bibinfo {volume} {63}},\ \bibinfo {pages} {5139}
  (\bibinfo {year} {1975})}\BibitemShut {NoStop}%
\bibitem [{\citenamefont {Blandin}\ \emph {et~al.}(1982)\citenamefont
  {Blandin}, \citenamefont {Larionov},\ and\ \citenamefont
  {Pasechnik}}]{82blandin}%
  \BibitemOpen
  \bibfield  {author} {\bibinfo {author} {\bibfnamefont {V.}~\bibnamefont
  {Blandin}}, \bibinfo {author} {\bibfnamefont {A.}~\bibnamefont {Larionov}}, \
  and\ \bibinfo {author} {\bibfnamefont {S.}~\bibnamefont {Pasechnik}},\
  }\href@noop {} {\bibfield  {journal} {\bibinfo  {journal} {J. Exp. Theor.
  Phys.}\ }\textbf {\bibinfo {volume} {56}},\ \bibinfo {pages} {1230} (\bibinfo
  {year} {1982})}\BibitemShut {NoStop}%
\bibitem [{\citenamefont {Oswald}(2015)}]{15Oswald}%
  \BibitemOpen
  \bibfield  {author} {\bibinfo {author} {\bibfnamefont {P.}~\bibnamefont
  {Oswald}},\ }\href {\doibase 10.1103/PhysRevE.92.062508} {\bibfield
  {journal} {\bibinfo  {journal} {Phys. Rev. E}\ }\textbf {\bibinfo {volume}
  {92}},\ \bibinfo {pages} {062508} (\bibinfo {year} {2015})}\BibitemShut
  {NoStop}%
\bibitem [{\citenamefont {Kahl}\ \emph {et~al.}(2013)\citenamefont {Kahl},
  \citenamefont {Baroni},\ and\ \citenamefont {Noirez}}]{13kahl}%
  \BibitemOpen
  \bibfield  {author} {\bibinfo {author} {\bibfnamefont {P.}~\bibnamefont
  {Kahl}}, \bibinfo {author} {\bibfnamefont {P.}~\bibnamefont {Baroni}}, \ and\
  \bibinfo {author} {\bibfnamefont {L.}~\bibnamefont {Noirez}},\ }\href
  {\doibase 10.1103/PhysRevE.88.050501} {\bibfield  {journal} {\bibinfo
  {journal} {Phys. Rev. E}\ }\textbf {\bibinfo {volume} {88}},\ \bibinfo
  {pages} {050501} (\bibinfo {year} {2013})}\BibitemShut {NoStop}%
\bibitem [{\citenamefont {Sellers}\ \emph {et~al.}(1988)\citenamefont
  {Sellers}, \citenamefont {Margulies},\ and\ \citenamefont
  {Schwarz}}]{88sellers}%
  \BibitemOpen
  \bibfield  {author} {\bibinfo {author} {\bibfnamefont {H.~S.}\ \bibnamefont
  {Sellers}}, \bibinfo {author} {\bibfnamefont {T.}~\bibnamefont {Margulies}},
  \ and\ \bibinfo {author} {\bibfnamefont {W.~H.}\ \bibnamefont {Schwarz}},\
  }\href@noop {} {\bibfield  {journal} {\bibinfo  {journal} {Mol. Cryst. Liq.
  Cryst.}\ }\textbf {\bibinfo {volume} {162}},\ \bibinfo {pages} {185}
  (\bibinfo {year} {1988})}\BibitemShut {NoStop}%
\bibitem [{\citenamefont {Biscari}\ \emph {et~al.}(2014)\citenamefont
  {Biscari}, \citenamefont {DiCarlo},\ and\ \citenamefont {Turzi}}]{14bdt}%
  \BibitemOpen
  \bibfield  {author} {\bibinfo {author} {\bibfnamefont {P.}~\bibnamefont
  {Biscari}}, \bibinfo {author} {\bibfnamefont {A.}~\bibnamefont {DiCarlo}}, \
  and\ \bibinfo {author} {\bibfnamefont {S.~S.}\ \bibnamefont {Turzi}},\ }\href
  {\doibase 10.1039/C4SM01067A} {\bibfield  {journal} {\bibinfo  {journal}
  {Soft Matter}\ }\textbf {\bibinfo {volume} {10}},\ \bibinfo {pages} {8296}
  (\bibinfo {year} {2014})}\BibitemShut {NoStop}%
\bibitem [{\citenamefont {Turzi}(2015)}]{15tur}%
  \BibitemOpen
  \bibfield  {author} {\bibinfo {author} {\bibfnamefont {S.~S.}\ \bibnamefont
  {Turzi}},\ }\href {\doibase 10.1017/S0956792514000345} {\bibfield  {journal}
  {\bibinfo  {journal} {Eur. J. Appl. Math.}\ }\textbf {\bibinfo {volume}
  {26}},\ \bibinfo {pages} {93} (\bibinfo {year} {2015})}\BibitemShut {NoStop}%
\bibitem [{\citenamefont {Biscari}\ \emph {et~al.}(2016)\citenamefont
  {Biscari}, \citenamefont {DiCarlo},\ and\ \citenamefont {Turzi}}]{16bdt}%
  \BibitemOpen
  \bibfield  {author} {\bibinfo {author} {\bibfnamefont {P.}~\bibnamefont
  {Biscari}}, \bibinfo {author} {\bibfnamefont {A.}~\bibnamefont {DiCarlo}}, \
  and\ \bibinfo {author} {\bibfnamefont {S.~S.}\ \bibnamefont {Turzi}},\ }\href
  {\doibase 10.1103/PhysRevE.93.052704} {\bibfield  {journal} {\bibinfo
  {journal} {Phys. Rev. E}\ }\textbf {\bibinfo {volume} {93}},\ \bibinfo
  {pages} {052704} (\bibinfo {year} {2016})}\BibitemShut {NoStop}%
\bibitem [{\citenamefont {Lord~Jr.}\ and\ \citenamefont
  {Labes}(1970)}]{70lord}%
  \BibitemOpen
  \bibfield  {author} {\bibinfo {author} {\bibfnamefont {A.~E.}\ \bibnamefont
  {Lord~Jr.}}\ and\ \bibinfo {author} {\bibfnamefont {M.~M.}\ \bibnamefont
  {Labes}},\ }\href@noop {} {\bibfield  {journal} {\bibinfo  {journal} {Phys.
  Rev. Lett.}\ }\textbf {\bibinfo {volume} {25}},\ \bibinfo {pages} {570}
  (\bibinfo {year} {1970})}\BibitemShut {NoStop}%
\bibitem [{\citenamefont {Kapustina}(2004)}]{Kapustina2004}%
  \BibitemOpen
  \bibfield  {author} {\bibinfo {author} {\bibfnamefont {O.~A.}\ \bibnamefont
  {Kapustina}},\ }\href {\doibase 10.1134/1.1780637} {\bibfield  {journal}
  {\bibinfo  {journal} {Crystallogr. Rep. +}\ }\textbf {\bibinfo {volume}
  {49}},\ \bibinfo {pages} {680} (\bibinfo {year} {2004})}\BibitemShut
  {NoStop}%
\bibitem [{\citenamefont {Kozhevnikov}(2005)}]{Kozhevnikov2005}%
  \BibitemOpen
  \bibfield  {author} {\bibinfo {author} {\bibfnamefont {E.~N.}\ \bibnamefont
  {Kozhevnikov}},\ }\href {\doibase 10.1134/1.2130900} {\bibfield  {journal}
  {\bibinfo  {journal} {Acoust. Phys. +}\ }\textbf {\bibinfo {volume} {51}},\
  \bibinfo {pages} {688} (\bibinfo {year} {2005})}\BibitemShut {NoStop}%
\bibitem [{\citenamefont {Kapustina}(2008)}]{Kapustina2008}%
  \BibitemOpen
  \bibfield  {author} {\bibinfo {author} {\bibfnamefont {O.~A.}\ \bibnamefont
  {Kapustina}},\ }\href {\doibase 10.1134/S1063771008060055} {\bibfield
  {journal} {\bibinfo  {journal} {Acoust. Phys. +}\ }\textbf {\bibinfo {volume}
  {54}},\ \bibinfo {pages} {778} (\bibinfo {year} {2008})}\BibitemShut
  {NoStop}%
\bibitem [{\citenamefont {Kozhevnikov}(2010)}]{Kozhevnikov2010}%
  \BibitemOpen
  \bibfield  {author} {\bibinfo {author} {\bibfnamefont {E.~N.}\ \bibnamefont
  {Kozhevnikov}},\ }\href {\doibase 10.1134/S1063771010010045} {\bibfield
  {journal} {\bibinfo  {journal} {Acoust. Phys. +}\ }\textbf {\bibinfo {volume}
  {56}},\ \bibinfo {pages} {24} (\bibinfo {year} {2010})}\BibitemShut {NoStop}%
\bibitem [{\citenamefont {Kapustina}\ \emph {et~al.}(2014)\citenamefont
  {Kapustina}, \citenamefont {Kozhevnikov},\ and\ \citenamefont
  {Chumakova}}]{Kapustina2014}%
  \BibitemOpen
  \bibfield  {author} {\bibinfo {author} {\bibfnamefont {O.~A.}\ \bibnamefont
  {Kapustina}}, \bibinfo {author} {\bibfnamefont {E.~N.}\ \bibnamefont
  {Kozhevnikov}}, \ and\ \bibinfo {author} {\bibfnamefont {S.~P.}\ \bibnamefont
  {Chumakova}},\ }\href {\doibase 10.1134/S1063771014030099} {\bibfield
  {journal} {\bibinfo  {journal} {Acoust. Phys. +}\ }\textbf {\bibinfo {volume}
  {60}},\ \bibinfo {pages} {269} (\bibinfo {year} {2014})}\BibitemShut
  {NoStop}%
\bibitem [{\citenamefont {Selinger}\ \emph {et~al.}(2002)\citenamefont
  {Selinger}, \citenamefont {Spector}, \citenamefont {Greanya}, \citenamefont
  {Weslowski}, \citenamefont {Shenoy},\ and\ \citenamefont
  {Shashidhar}}]{02seli}%
  \BibitemOpen
  \bibfield  {author} {\bibinfo {author} {\bibfnamefont {J.~V.}\ \bibnamefont
  {Selinger}}, \bibinfo {author} {\bibfnamefont {M.~S.}\ \bibnamefont
  {Spector}}, \bibinfo {author} {\bibfnamefont {V.~A.}\ \bibnamefont
  {Greanya}}, \bibinfo {author} {\bibfnamefont {B.}~\bibnamefont {Weslowski}},
  \bibinfo {author} {\bibfnamefont {D.}~\bibnamefont {Shenoy}}, \ and\ \bibinfo
  {author} {\bibfnamefont {R.}~\bibnamefont {Shashidhar}},\ }\href@noop {}
  {\bibfield  {journal} {\bibinfo  {journal} {Phys. Rev. E}\ }\textbf {\bibinfo
  {volume} {66}},\ \bibinfo {pages} {051708} (\bibinfo {year}
  {2002})}\BibitemShut {NoStop}%
\bibitem [{\citenamefont {Greanya}\ \emph {et~al.}(2003)\citenamefont
  {Greanya}, \citenamefont {Spector}, \citenamefont {Selinger}, \citenamefont
  {Weslowski},\ and\ \citenamefont {Shashidhar}}]{03seli}%
  \BibitemOpen
  \bibfield  {author} {\bibinfo {author} {\bibfnamefont {V.~A.}\ \bibnamefont
  {Greanya}}, \bibinfo {author} {\bibfnamefont {M.~S.}\ \bibnamefont
  {Spector}}, \bibinfo {author} {\bibfnamefont {J.~V.}\ \bibnamefont
  {Selinger}}, \bibinfo {author} {\bibfnamefont {B.~T.}\ \bibnamefont
  {Weslowski}}, \ and\ \bibinfo {author} {\bibfnamefont {R.}~\bibnamefont
  {Shashidhar}},\ }\href@noop {} {\bibfield  {journal} {\bibinfo  {journal}
  {J.\ Appl.\ Phys.}\ }\textbf {\bibinfo {volume} {94}},\ \bibinfo {pages}
  {7571} (\bibinfo {year} {2003})}\BibitemShut {NoStop}%
\bibitem [{\citenamefont {Malanoski}\ \emph {et~al.}(2004)\citenamefont
  {Malanoski}, \citenamefont {Greanya}, \citenamefont {Weslowski},
  \citenamefont {Spector}, \citenamefont {Selinger},\ and\ \citenamefont
  {Shashidhar}}]{04seli}%
  \BibitemOpen
  \bibfield  {author} {\bibinfo {author} {\bibfnamefont {A.~P.}\ \bibnamefont
  {Malanoski}}, \bibinfo {author} {\bibfnamefont {V.~A.}\ \bibnamefont
  {Greanya}}, \bibinfo {author} {\bibfnamefont {B.~T.}\ \bibnamefont
  {Weslowski}}, \bibinfo {author} {\bibfnamefont {M.~S.}\ \bibnamefont
  {Spector}}, \bibinfo {author} {\bibfnamefont {J.~V.}\ \bibnamefont
  {Selinger}}, \ and\ \bibinfo {author} {\bibfnamefont {R.}~\bibnamefont
  {Shashidhar}},\ }\href@noop {} {\bibfield  {journal} {\bibinfo  {journal}
  {Phys. Rev. E}\ }\textbf {\bibinfo {volume} {69}},\ \bibinfo {pages} {021705}
  (\bibinfo {year} {2004})}\BibitemShut {NoStop}%
\bibitem [{\citenamefont {Greanya}\ \emph {et~al.}(2005)\citenamefont
  {Greanya}, \citenamefont {Malanoski}, \citenamefont {Weslowski},
  \citenamefont {Spector},\ and\ \citenamefont {Selinger}}]{05seli}%
  \BibitemOpen
  \bibfield  {author} {\bibinfo {author} {\bibfnamefont {V.~A.}\ \bibnamefont
  {Greanya}}, \bibinfo {author} {\bibfnamefont {A.~P.}\ \bibnamefont
  {Malanoski}}, \bibinfo {author} {\bibfnamefont {B.~T.}\ \bibnamefont
  {Weslowski}}, \bibinfo {author} {\bibfnamefont {M.~S.}\ \bibnamefont
  {Spector}}, \ and\ \bibinfo {author} {\bibfnamefont {J.~V.}\ \bibnamefont
  {Selinger}},\ }\href@noop {} {\bibfield  {journal} {\bibinfo  {journal}
  {Liq.\ Cryst.}\ }\textbf {\bibinfo {volume} {32}},\ \bibinfo {pages} {933}
  (\bibinfo {year} {2005})}\BibitemShut {NoStop}%
\bibitem [{\citenamefont {Virga}(2009)}]{09virga}%
  \BibitemOpen
  \bibfield  {author} {\bibinfo {author} {\bibfnamefont {E.~G.}\ \bibnamefont
  {Virga}},\ }\href {\doibase 10.1103/PhysRevE.80.031705} {\bibfield  {journal}
  {\bibinfo  {journal} {Phys. Rev. E}\ }\textbf {\bibinfo {volume} {80}},\
  \bibinfo {pages} {031705} (\bibinfo {year} {2009})}\BibitemShut {NoStop}%
\bibitem [{\citenamefont {DiCarlo}\ and\ \citenamefont
  {Quiligotti}(2002)}]{02DiCarlo}%
  \BibitemOpen
  \bibfield  {author} {\bibinfo {author} {\bibfnamefont {A.}~\bibnamefont
  {DiCarlo}}\ and\ \bibinfo {author} {\bibfnamefont {S.}~\bibnamefont
  {Quiligotti}},\ }\href {\doibase
  http://dx.doi.org/10.1016/S0093-6413(02)00297-5} {\bibfield  {journal}
  {\bibinfo  {journal} {Mech. Res. Commun.}\ }\textbf {\bibinfo {volume}
  {29}},\ \bibinfo {pages} {449} (\bibinfo {year} {2002})}\BibitemShut
  {NoStop}%
\bibitem [{\citenamefont {Rajagopal}\ and\ \citenamefont
  {Srinivasa}(2004)}]{04Rajagopal}%
  \BibitemOpen
  \bibfield  {author} {\bibinfo {author} {\bibfnamefont {K.~R.}\ \bibnamefont
  {Rajagopal}}\ and\ \bibinfo {author} {\bibfnamefont {A.~R.}\ \bibnamefont
  {Srinivasa}},\ }\href {\doibase 10.1007/s00033-004-4019-6} {\bibfield
  {journal} {\bibinfo  {journal} {Z. Angew. Math. Phys.}\ }\textbf {\bibinfo
  {volume} {55}},\ \bibinfo {pages} {861} (\bibinfo {year} {2004})}\BibitemShut
  {NoStop}%
\bibitem [{\citenamefont {Podio-Guidugli}\ and\ \citenamefont
  {Virga}(1987)}]{87povi}%
  \BibitemOpen
  \bibfield  {author} {\bibinfo {author} {\bibfnamefont {P.}~\bibnamefont
  {Podio-Guidugli}}\ and\ \bibinfo {author} {\bibfnamefont {E.~G.}\
  \bibnamefont {Virga}},\ }\href {\doibase 10.1098/rspa.1987.0055} {\bibfield
  {journal} {\bibinfo  {journal} {Proc. R. Soc. London, Ser. A}\ }\textbf
  {\bibinfo {volume} {411}},\ \bibinfo {pages} {85} (\bibinfo {year}
  {1987})}\BibitemShut {NoStop}%
\bibitem [{Note1()}]{Note1}%
  \BibitemOpen
  \bibinfo {note} {The scalar product $\left < \cdot , \cdot \right >$ is well
  defined when $a(\rho )>0$ since the tensor $\protect \bm {\Psi }^{-1}
  \protect \tp \protect \bm {\Psi }^{-1}$ is then positive definite and
  symmetric}\BibitemShut {NoStop}%
\bibitem [{\citenamefont {Parodi}(1970)}]{70paro}%
  \BibitemOpen
  \bibfield  {author} {\bibinfo {author} {\bibfnamefont {O.}~\bibnamefont
  {Parodi}},\ }\href {\doibase 10.1051/jphys:01970003107058100} {\bibfield
  {journal} {\bibinfo  {journal} {J. Phys. France}\ }\textbf {\bibinfo {volume}
  {31}},\ \bibinfo {pages} {581} (\bibinfo {year} {1970})}\BibitemShut
  {NoStop}%
\bibitem [{\citenamefont {Fedorov}(1968)}]{68Fedorov}%
  \BibitemOpen
  \bibfield  {author} {\bibinfo {author} {\bibfnamefont {F.}~\bibnamefont
  {Fedorov}},\ }\href@noop {} {\emph {\bibinfo {title} {Theory of elastic waves
  in crystals}}}\ (\bibinfo  {publisher} {Plenum Press},\ \bibinfo {address}
  {New York},\ \bibinfo {year} {1968})\BibitemShut {NoStop}%
\bibitem [{\citenamefont {Bird}\ \emph {et~al.}(1987)\citenamefont {Bird},
  \citenamefont {Armstrong},\ and\ \citenamefont {Hassager}}]{87bird}%
  \BibitemOpen
  \bibfield  {author} {\bibinfo {author} {\bibfnamefont {R.~B.}\ \bibnamefont
  {Bird}}, \bibinfo {author} {\bibfnamefont {R.~C.}\ \bibnamefont {Armstrong}},
  \ and\ \bibinfo {author} {\bibfnamefont {O.}~\bibnamefont {Hassager}},\
  }\href@noop {} {\emph {\bibinfo {title} {Dynamics of Polymeric Liquids,
  Volume 1: Fluid Mechanics}}},\ \bibinfo {edition} {2nd}\ ed.\ (\bibinfo
  {publisher} {Wiley},\ \bibinfo {year} {1987})\BibitemShut {NoStop}%
\bibitem [{\citenamefont {De~Matteis}\ and\ \citenamefont
  {Virga}(2011)}]{11DeMatteis}%
  \BibitemOpen
  \bibfield  {author} {\bibinfo {author} {\bibfnamefont {G.}~\bibnamefont
  {De~Matteis}}\ and\ \bibinfo {author} {\bibfnamefont {E.~G.}\ \bibnamefont
  {Virga}},\ }\href {\doibase 10.1103/PhysRevE.83.011703} {\bibfield  {journal}
  {\bibinfo  {journal} {Phys. Rev. E}\ }\textbf {\bibinfo {volume} {83}},\
  \bibinfo {pages} {011703} (\bibinfo {year} {2011})}\BibitemShut {NoStop}%
\bibitem [{\citenamefont {Kneppe}\ \emph {et~al.}(1982)\citenamefont {Kneppe},
  \citenamefont {Schneider},\ and\ \citenamefont {Sharma}}]{82kneppe}%
  \BibitemOpen
  \bibfield  {author} {\bibinfo {author} {\bibfnamefont {H.}~\bibnamefont
  {Kneppe}}, \bibinfo {author} {\bibfnamefont {F.}~\bibnamefont {Schneider}}, \
  and\ \bibinfo {author} {\bibfnamefont {N.~K.}\ \bibnamefont {Sharma}},\
  }\href@noop {} {\bibfield  {journal} {\bibinfo  {journal} {J. Chem. Phys.}\
  }\textbf {\bibinfo {volume} {77}},\ \bibinfo {pages} {3203} (\bibinfo {year}
  {1982})}\BibitemShut {NoStop}%
\end{thebibliography}

%

\end{document}